\documentstyle[aps,prb,epsf]{revtex}

\newcommand{\bfc}{{\bf c}}
\newcommand{\bfx}{{\bf x}}
\newcommand{\bfy}{{\bf y}}

\newcommand{\dt}{\Delta t}
\newcommand{\Schrodinger}{Schr\"{o}dinger}
\newcommand{\tpdt}{t + \dt}

\newcommand{\diag}{\mbox{diag}}

\begin{document}

\title{
\begin{flushleft}
{\small BU-CCS-960401}\\
{\small PUPT-1615}\\
{\small quant-ph/9604035}\\
\end{flushleft}
A Quantum Lattice-Gas Model for\\
the Many-Particle \Schrodinger\ Equation\\
in $d$ Dimensions
}
\author{
Bruce M. Boghosian\\
{\small \sl Center for Computational Science,} \\
{\small \sl Boston University,} \\
{\small \sl 3 Cummington Street, Boston, Massachusetts 02215, U.S.A.} \\
{\small \tt bruceb@bu.edu} \\[0.3cm]
Washington Taylor IV\\
{\small \sl Department of Physics,} \\
{\small \sl Princeton University,} \\
{\small \sl Princeton, New Jersey 08544, U.S.A.} \\
{\small \tt wati@princeton.edu} \\[0.3cm]
}
\date{\today}
\maketitle

\begin{abstract}
  We consider a general class of discrete unitary dynamical models on
  the lattice.  We show that generically such models give rise to a
  wavefunction satisfying a \Schrodinger\ equation in the continuum
  limit, in any number of dimensions.  There is a simple mathematical
  relationship between the mass of the \Schrodinger\ particle and the
  eigenvalues of a unitary matrix describing the local evolution of
  the model.  Second quantized versions of these unitary models can be
  defined, describing in the continuum limit the evolution of a
  nonrelativistic quantum many-body theory.  An arbitrary potential is
  easily incorporated into these systems.  The models we describe fall
  in the class of quantum lattice gas automata, and can be implemented
  on a quantum computer with a speedup exponential in the number of
  particles in the system.  This gives an efficient algorithm for
  simulating general nonrelativistic interacting quantum many-body
  systems on a quantum computer.
\end{abstract}

\section{Introduction}
\label{sec:intro}

There are many situations in physics where a continuous system obeying
a particular set of equations at a macroscopic scale can be modeled by
a discrete microscopic system obeying a very simple set of local
rules.  For example, in equilibrium statistical mechanics, simple
lattice models such as the Ising model capture the behavior of generic
classes of critical systems at large scales.  Another interesting
class of discrete systems are lattice gas
automata~\cite{bib:fhp,bib:fchc,bib:swolf}; these models describe
systems of particles moving about on a lattice, obeying simple
collision rules which conserve quantities such as mass and momentum.  In
the macroscopic limit, these systems obey Navier-Stokes or other
hydrodynamic equations of interest.

In the quantum domain, there are also examples of discrete microscopic
systems which capture interesting macroscopic behavior.  Lattice-gauge
theories (see for example Creutz~\cite{bib:latticegauge}) give an
approach to studying the partition function and spectra of quantum field
theories by mapping these theories to statistical mechanical ensembles.
There are, however, few discrete models for describing the dynamical
evolution of quantum systems which preserve important features such as
unitarity.  An example of a quantum system for which a unitary discrete
model is known is the Dirac equation describing a relativistic particle
moving in one spatial dimension.  As shown by
Feynman~\cite{bib:feynsch,bib:feynhibbs}, this system can be described
by a simple microscopic model of a particle moving on a 1D lattice
according to a simple local rule which essentially corresponds to a
unitary form of random walk.  A straightforward attempt to realize a
Dirac equation in more than one spatial dimension as a form of unitary
random walk cannot succeed.  By using operator splitting methods,
however, it was shown by Succi and Benzi~\cite{bib:sb} that a sequence
of random moves along single axes, alternating with transformations
which diagonalize each of the Dirac matrices in turn, can give an
analogous construction in higher dimensions.  The discrete model for the
(1 + 1)-dimensional Dirac equation has been of renewed interest
recently\cite{bib:sb,bib:kn,bib:meyer}, due partly to the possibility of
simulating such unitary microscopic discrete systems by quantum
computers.  In particular, it has recently been
suggested\cite{bib:meyer} that a simple quantum lattice model can be
constructed which describes the motion of a system of many particles
moving according to the one dimensional Dirac equation.

In this paper we consider a class of models closely related to the
1D Dirac lattice model, which give rise to a nonrelativistic
single-particle \Schrodinger\ equation in an arbitrary number of
dimensions.  In these models, the time development rule is given by a
single local, unitary transformation matrix.  Thus, we are essentially
considering the motion of a single particle under a unitary random walk
process.  For this class of models we show that the macroscopic equation
of motion satisfied by the wavefunction corresponding to particle
density is the \Schrodinger\ equation.  We show that such
nonrelativistic models can be constructed for an arbitrary number of
spatial dimensions.  We also show that an arbitrary potential can easily
be included in these models.

It is natural to generalize from the single particle models to a second
quantized many-body system.  Such a model could be implemented very
efficiently on a quantum computer, so that the number of computational
steps necessary to simulate a single time step would only depend upon
the size of the lattice and would not depend upon the number of
particles in the system being simulated.  Thus, our results could be
used to efficiently simulate an arbitrary nonrelativistic interacting
quantum many-body system on a quantum computer exponentially faster
than the same calculation could be performed on a classical computer.
Such a system would be an ideal example of quantum computing, since
the computing elements could be built from a system of spin-$1/2$
particles on a lattice obeying simple local unitary time evolution
rules.

The principal difference between our models and the 1D Dirac lattice
model (and its generalization by Succi and Benzi~\cite{bib:sb}) is
that in the Dirac model, the unitary evolution rule satisfied by the
wave function or particle at each time step is infinitesimally close
to the identity transformation.  As the lattice spacing $\epsilon$
goes to 0, the unitary transition matrix is of the form $S = 1 + i
\epsilon M$, where $M$ is Hermitian.  In our models, we take the
transition matrix to be independent of the lattice spacing.  This form
of a time development equation makes the system nonrelativistic, but
allows for a formulation in an arbitrary number of dimensions.  A
closely related model was considered recently in one spatial dimension
\cite{bib:succi}, where simulations were shown to be consistent with
emergent behavior corresponding to a \Schrodinger\ equation.  In this
paper we prove that this is the general behavior of such models,
giving a simple algebraic relation between the transition matrix and
the mass of the nonrelativistic particle.  We develop such models for
the \Schrodinger\ equation in an arbitrary number of dimensions.

In the first part of this paper we will consider lattice models for
single-particle motion.  These models are essentially unitary
lattice-Boltzmann models\cite{bib:lb}.  In the latter part of the
paper we generalize to many-body systems and discuss how a lattice of
quantum computing elements could be used to describe the motion of a
large number of nonrelativistic quantum particles.  We conclude with a
simple numerical check of the analytic description of a sample model.

As a simple example of the type of system considered in this paper,
consider the lattice-Boltzmann model with a configuration space defined
by two complex fields, $\psi_1 (x, t)$ and $\psi_2 (x, t)$, taking
independent values on a lattice with one spatial dimension $x$ and one
temporal dimension $t$.  Define the dynamics of this model to obey the
equations
\[
\psi_1 (x + 1, t) = \frac{1}{ 2}
\left[ (1- i) \psi_1 (x, t -1)  -(1 + i) \psi_2 (x, t -1) \right]
\]
\[
\psi_2 (x - 1, t) = \frac{1}{ 2}
\left[  (1- i)\psi_2 (x, t -1)  -(1 + i) \psi_1 (x, t -1) \right].
\]
These equations give a unitary time evolution to $\psi$.  To understand
how $\psi$ evolves in a continuum limit, we can expand the equations of
motion through 4 time steps, giving for example
\begin{eqnarray*}
\psi_1 ( x, t + 4) & = & \frac{1}{4}  \left[
-\psi_1 (x-4,t)   +3 \psi_1 (x-2,t) +\psi_1 (x,t)  +
\psi_1 (x+2,t)\right]\\
 & & \hspace{0.3in}
+ \frac{i}{4}  \left[
\psi_2 (x-2,t) -\psi_2 (x,t)  -
\psi_2 (x+2,t) +\psi_2 (x+4,t) \right].
\end{eqnarray*}
Taking a continuous limit as the lattice spacing scales as $\epsilon$
in the $x$ direction and $\epsilon^2$ in the $t$ direction, we find
the differential equation
\[
 \partial_t {\psi_1} (t) =   \frac{i}{2}  \partial_x^2 \psi_2 (t)
\]
A similar equation holds for $\psi_2$, and so it follows that
\[
\partial_t (\psi_1 + \psi_2) =
  \frac{i}{2}  \partial_x^2 (\psi_1 + \psi_2).
\]
Thus, we see that the total amplitude $\Psi (x, t)= \psi_1 (x, t) +
\psi_2 (x, t)$ satisfies a \Schrodinger\ equation.  As we shall
demonstrate, this is the generic behavior of a unitary Boltzmann model
with a fixed time development rule.

We introduce the \Schrodinger\ model in Sec.~\ref{sec:se1d} by
presenting the one-dimensional case.  The model is generalized to
Cartesian lattices of arbitrary dimension in Sec.~\ref{sec:send}; in
this section we also discuss the inclusion of a potential.  In
Sec.~\ref{sec:gas}, we discuss how the one-particle models can be
generalized to construct a quantum lattice-gas model of many
nonrelativistic particles.  In Sec.~\ref{sec:numeric} we give the
results of a simulation of a single free nonrelativistic particle in 2D,
comparing numerical results with the theoretical framework presented
here.  The appendices give explicit formulas for models in 2D and 3D on
a Cartesian lattice.

\section{Schr\"{o}dinger equation in one dimension}
\label{sec:se1d}

In this section, we consider unitary lattice-Boltzmann models describing
the evolution of a single particle in one dimension.  Keeping the
collision operator fixed in the scaling limit, we show that a very
general class of microscopic models give rise in the continuum limit to
a \Schrodinger\ equation.

We define the model on a lattice given by points $x = \epsilon n$
where $n$ is an integer.  The lattice can be taken to either have
periodic boundary conditions or to be of infinite extent.  The state
of the system at a fixed value of the time parameter $t$ is described
by a wave function $\psi_k (x, t)$ which depends upon the discrete
position $x$ and an ``internal'' index $k$ taking values from 1 to
$m$, labeling possible particle velocities at the lattice site $x$.
As in lattice-Boltzmann models, at each discrete time step the various
components of the field at each site undergo a local unitary
``collision,'' and then the $j$th component of $\psi (x, t)$
propagates along the $j$th lattice vector $c_j$ to the new site $x +
c_j$ to yield the new state of the system at time $\tpdt$.  We
consider only linear processes, so this interaction can be specified
by an $m$-by-$m$ scattering matrix $S$.

We take the continuum limit of the theory by scaling $\epsilon
\rightarrow 0$, where $\Delta t \sim \epsilon^2$.  In this limit we
will find that the discrete equation describing the dynamics of $\psi$
becomes a continuous differential equation, which we identify as the
\Schrodinger\ equation.

In this section we will assume that each lattice site has two associated
possible particle velocities, corresponding to right- and left-moving
particles.  We will also assume that the dynamics is symmetric under
right-left reflection.  More general models can easily be analyzed using
a similar formalism.

The equation of motion for the model reads
\[
\psi_k  (x + \epsilon c_k, t) = S_{kj} \psi_j (x, t - \dt)
\]
where $k = 1, 2$ correspond to right and left moving particles, so that
$c_1 = +1, c_2 = -1$.  The quantum wave function $\psi_k (x, t)$ is
normalized so that
\begin{equation}
\sum_{x, k}| \psi_k (x, t) |^2 = 1
\label{eq:normalization}
\end{equation}
for all $t$.  The matrix $S_{kj}$ is a two-by-two matrix which is
unitary so as to preserve the condition Eq.~(\ref{eq:normalization}).  

We begin our analysis by transforming to the wavefunction
\[
\psi (x,t) = S^\tau \phi (x,t),
\]
where $\tau\equiv t/\dt$.  We can then expand the difference $\phi
(t)-\phi (t -1)$ in the infinitesimal parameter $\epsilon$, to get
\[
\phi (t)-\phi (t-1) =-\epsilon S^{-\tau} C S^\tau \partial_x \phi
- \frac{ \epsilon^2}{2} S^{-\tau}C^2
 S^\tau \partial_x^2 \phi+{\cal O} (\epsilon^3)
\]
where $C$ is the two-by-two matrix
\[
C =   \left(\begin{array}{cc}
1 & 0\\
0 & -1
\end{array} \right)
\]

Because we are assuming that the interaction described by the matrix
$S$ is invariant under reflection, $S$ must be of the form
\[
S =   \left(\begin{array}{cc}
a & b\\
b & a
\end{array} \right)
\]
where $a$ and $b$ are complex numbers.  Because of unitarity we have
$| a |^2+ | b |^2 = 1$.  $S$ can be put in diagonal form by writing
\[
S = X^{-1} D X
\]
where
\[
X = \frac{1}{ \sqrt{2}}  \left( \begin{array}{cc}
1 & 1\\
1 & -1
\end{array} \right).
\]
We can redefine $S$ and $\psi$ up to a phase, so that without loss of
generality we can take
\[
D =  \left( \begin{array}{cc}
\mu & 0\\
0 & 1
\end{array} \right).
\]
where $\mu$ is a complex number with magnitude 1 ($\mu \mu^*= 1$).
We then have
\[
S =  \frac{1}{2} \left( \begin{array}{cc}
\mu + 1 & \mu -1\\
\mu -1 & \mu +  1
\end{array} \right).
\]
If we write $X \phi= \eta$, then we have
\[
\eta (t)-\eta (t-1) = -\epsilon 
D^{-\tau} (X C X^{-1}) D^\tau \partial_x\eta
- \frac{ \epsilon^2}{2} D^{-\tau} (X C^2
X^{-1}) D^\tau \partial_x^2 \eta +{\cal O} (\epsilon^3)
\]

At this point we would like to scale the time step as a power of
$\epsilon$ so that this equation can be written as a differential
equation in time.  However, there is a difficulty which arises due to
the fact that there are two relevant time scales involved in the
dynamics of $\eta$.  There is an order-$\epsilon$ change to $\eta$ at
every time step; however, this order-$\epsilon$ term has a phase angle
which rotates at every time step.  Thus, the order-$\epsilon$ dynamics
average out after a large number of time steps, so that the
time-averaged rate of change of $\eta$ actually goes as $\epsilon^2$.
The dynamics we are interested in are independent of the short-term
order-$\epsilon$ fluctuations, so we must perform another transformation
to remove these effects.  With this goal in mind, we write
\[
\eta (t) = \zeta (t) + \epsilon \rho (t)
\]
where 
\[
\rho (t)-\rho (t-1) =
-D^{-\tau} (X C X^{-1}) D^\tau \partial_x\zeta.
\]
This equation is solved by
\[
\rho (t) = D^{-\tau} GD^\tau\partial_x \zeta,
\]
where
\[
G-DGD^{-1} = -B = -(X C X^{-1}).
\]
This can be solved for $G$ as long as the only nonzero entries
$B_{ij}$ appear where the $i$ and $j$ eigenvalues of $D$ are
different.  We can now write a final dynamical equation for $\zeta$.
\[
\zeta (t)-\zeta (t-1) = -\epsilon^2 D^{-\tau} B G D^\tau \partial_x^2
\zeta - \frac{ \epsilon^2}{2} D^{-\tau} (X C^2
X^{-1}) D^\tau \partial_x^2 \zeta +{\cal O} (\epsilon^3).
\]
If we assume that the unit of time scales as $\epsilon^2$, we have
the continuous dynamical equation
\[
\partial_t {\zeta} = - D^{-\tau} B G D^\tau \partial_x^2 \zeta
- \frac{1}{2} D^{-\tau} (X C^2
X^{-1}) D^\tau \partial_x^2 \zeta.
\]

We can now substitute the known matrices $X, C, D$ to compute
\[
B =X C X^{-1}= \left(\begin{array}{cc}
0 & 1\\
1 & 0
\end{array} \right)
\]
\[
X C^2 X^{-1}= \left(\begin{array}{cc}
1 & 0\\
0 & 1
\end{array} \right)
\]
\[
G =\left(\begin{array}{cc}
0 &  \frac{-1}{1-\mu} \\
\frac{-1}{1-\mu^*}  & 0
\end{array} \right).
\]
Using these matrices we have
\[
BG + \frac{1}{2}  XC^2 X^{-1}
= \left(\begin{array}{cc}
\frac{1}{2} - \frac{1}{1 -\mu^*}   & 0\\
0 & \frac{1}{2} - \frac{1}{1 -\mu}
\end{array} \right)
\]
Writing $\mu = \cos \theta + i \sin \theta$, we have
\[
\frac{1}{1 -\mu}
= \frac{1-\cos \theta + i \sin \theta}{ (1 -\cos \theta)^2+ \sin^2
\theta} 
= \frac{1}{2} + i \frac{\sin \theta}{2 (1 -\cos \theta)}.
\]
Thus, the dynamical equation for $\zeta$ becomes
\[
\partial_t {\zeta} = i 
\left(\begin{array}{cc}
\frac{1}{2 m} & 0 \\
0 &-\frac{1}{2 m}
\end{array} \right) \partial_x^2 \zeta
\]
where 
\[
m =\cot \theta-\csc \theta.
\]
The equation for the first component of $\zeta$ is thus precisely a
\Schrodinger\ equation for a particle moving in one dimension with mass
$m$.  To leading order, $\zeta (t)$ is related to $\psi$ through the
sequence of transformations described above, so that
\[
\zeta (t) = D^{-\tau} X \psi (t) +{\cal O} (\epsilon)
\]
The first component of $\zeta (t)$ is therefore given by
\[
\Psi =
\zeta_1 (t) = \frac{\mu^{-\tau}}{\sqrt{2}}  \left(\psi_1 (t) + \psi_2
(t)\right), 
\]
and this satisfies the \Schrodinger\ equation in the continuum limit,
\[
\partial_t {\Psi} = i \frac{1}{2 m}  \partial_x^2 \Psi.
\]
Note that by taking $ \mu = -i$ we get $m = 1$, giving precisely the
example discussed in Sec.~\ref{sec:intro}.  We shall demonstrate in
Sec.~\ref{sec:send} that, in an analogous fashion, in higher-dimensional
theories the sum of the wave function components forms a scalar quantity
which satisfies a \Schrodinger\ equation.

\section{Schr\"{o}dinger equation in dimensions $d \geq 1$}
\label{sec:send}

In this section we derive the general form for the continuum limit of
the dynamics for a unitary lattice-Boltzmann model with fixed collision
matrix on a lattice with any number of dimensions.  Specializing to the
case where the lattice is Cartesian and the collision rule is invariant
under discrete rotations, we find that a generic collision rule gives a
\Schrodinger\ equation in any dimension $d$.

\subsection{General form of dynamical equation}
The analysis of the continuous equations of motion in $d$ dimensions
proceeds in a fashion very similar to the discussion in the previous
section.  We assume that the lattice contains a set of points $\bfx$,
and that at each lattice site there are particle velocities labeled by
$k$, corresponding to velocity vectors $\bfc_k$ in the lattice.
Denoting spatial indices by $\alpha$, we denote the $\alpha$th component
of the $k$th velocity vector by $c_k^\alpha$.  The dynamics of the
lattice-Boltzmann model are described by the equation of motion
\[
\psi_k  (\bfx + \epsilon\bfc_k, t) = S_{kj} \psi_j (\bfx, t -1)
\]
where $S$ is unitary.
Transforming as before
\[
\psi (t) = S^\tau \phi (t)
\]
we have
\[
\phi (t)-\phi (t -1) =
-\epsilon S^{-\tau} C^{\alpha} S^\tau \partial_\alpha \phi
- \frac{ \epsilon^2}{2} S^{-\tau}C^{\alpha} 
C^{\beta} S^\tau \partial_\alpha \partial_\beta \phi
+{\cal O} (\epsilon^3)
\]
where the diagonal matrices $C^\alpha$ are given by
\[
  C^\alpha\equiv
  \diag\left(c^\alpha_1,\ldots,c^\alpha_n\right),
\]
with $c^\alpha_j$ being the $\alpha$th spatial component of the $j$th
lattice vector.  Writing $S = X^{-1} D X$, $X\phi = \eta$ we have
\[
\eta (t)-\eta (t-1) = -\epsilon 
D^{-\tau} (X C^{\alpha} X^{-1}) D^\tau \partial_\alpha\eta
- \frac{ \epsilon^2}{2} D^{-\tau} (X C^{\alpha} C^{\beta} 
X^{-1}) D^\tau \partial_\alpha
\partial_\beta \eta+{\cal O} (\epsilon^3)
\]
We write
\[
\eta (t) = \zeta (t) + \epsilon \rho (t)
\]
where
\[
\rho (t)-\rho (t-1) =
-D^{-\tau} (X C^{\alpha} X^{-1}) D^\tau \partial_\alpha\zeta
\]
This is solved, as before, by
\[
\rho (t) = D^{-\tau} G^{\alpha}D^\tau\partial_\alpha \zeta,
\]
where
\[
G-DGD^{-1} = -(X C X^{-1}) = -B
\]
Again, this can be solved for $G$ as long as the only nonzero entries
$B_{ij}$ appear where the $i$ and $j$ eigenvalues of $D$ are
different.  The resulting continuum equation for $\eta$ is
\[
\partial_t {\eta} = - D^{-\tau} B^\alpha G^\beta D^\tau \partial_\alpha
\partial_\beta \eta
- \frac{1}{2} D^{-\tau} (X C^{\alpha} C^{\beta} 
X^{-1}) D^\tau \partial_\alpha
\partial_\beta \eta
\]
This is the general form of the dynamical equation for a unitary
lattice-Boltzmann model.

\subsection{Schr\"{o}dinger equation in $d$ dimensions}

We now specialize to the case where the lattice is Cartesian, so that
there are $2d$ possible particle velocities at each lattice site,
corresponding to vectors of magnitude $+ \epsilon, -\epsilon$ in each of
the $d$ directions.  We choose the collision matrix $S$ to be invariant
under the symmetry group of the lattice.  We will show that generically
the continuum limit of the equation of motion is a \Schrodinger\ 
equation, just as we found for a general collision matrix in 1D on the
Cartesian lattice.

The constraint that $S$ is invariant under discrete rotations and
reflections is actually quite a strong condition.  The $2d$-dimensional
space of velocity vectors transforms under a linear representation of
this discrete group.  This representation contains only 3 irreducible
representations, which allows us to determine $D$ up to 3 distinct
eigenvalues.  Because of the symmetry constraint, we can always
diagonalize $S$ by the matrix
\[
X = \left(\begin{array}{ccccccccccc}
\frac{1}{ \sqrt{2 d}}  & \frac{1}{ \sqrt{2 d}}  & \frac{1}{ \sqrt{2 d}}  &
\cdots & \frac{1}{ \sqrt{2 d}}  & \frac{1}{ \sqrt{2 d}}  
& \frac{1}{ \sqrt{2 d}}  & \frac{1}{
\sqrt{2 d}}  & \cdots & \frac{1}{ \sqrt{2 d}}  & \frac{1}{ \sqrt{2 d}} \\ 
\frac{1}{ \sqrt{2}}  & 0 & 0 & \cdots & 0 & - \frac{1}{ \sqrt{2}}  & 0 &
0 & \cdots & 0 & 0\\
0 & \frac{1}{ \sqrt{2}}  & 0 & \cdots & 0 & 0 &- \frac{1}{ \sqrt{2}} &
0 & \cdots & 0 & 0\\
\vdots & \vdots & \vdots &  & \vdots
 & \vdots & \vdots & \vdots & & \vdots & \vdots\\
0 & 0 & 0 & \cdots & \frac{1}{ \sqrt{2}}  & 0 & 0
 & 0 & \cdots & 0 & -\frac{1}{ \sqrt{2}} \\
\frac{1}{ 2}  & -\frac{1}{ 2}  & 0 & \cdots & 0 &
\frac{1}{ 2}  & -\frac{1}{ 2} & 0 & \cdots & 0 & 0\\
\frac{1}{ 2 \sqrt{3}}  & \frac{1}{ 2 \sqrt{3}}  & -\frac{2}{ 2 \sqrt{3}}
 & \cdots & 0 & \frac{1}{ 2 \sqrt{3}}  & \frac{1}{ 2 \sqrt{3}} &
-\frac{2}{ 2 \sqrt{3}}  & \cdots &0 & 0\\
\vdots & \vdots & \vdots &  & \vdots
 & \vdots & \vdots &\vdots & & \vdots & \vdots\\
\frac{1}{ 2 \sqrt{d_2}} & \frac{1}{ 2 \sqrt{d_2}} & \frac{1}{ 2 \sqrt{d_2}} &
\cdots & \frac{1 -d}{ 2 \sqrt{d_2}} & \frac{1}{ 2 \sqrt{d_2}}  
& \frac{1}{ 2 \sqrt{d_2}}  & \frac{1}{
2 \sqrt{d_2}}  & \cdots & \frac{1}{ 2 \sqrt{d_2}}  & \frac{1
-d}{ 2 \sqrt{d_2}}\\ 
\end{array} \right)
\]
where $d_2 = d (d-1)/2$.  The rows of this matrix consist of the 3
groups of vectors in the irreducible representations of the rotation
group mentioned above.  The first row is the normalized vector $(1,
1,\ldots,1)$.  The next $d$ rows are normalized versions of the
vectors $\bfc^\alpha$ with $+1$ in position $i$ and $-1$ in position $i +
d$.  The last $d-1$ rows are vectors with equal components $i$ and $i
+ d$, subject to the condition that the sum of the components
vanishes.  This matrix puts $S$ in the diagonal form
\[
D = XSX^{-1}  = \left(\begin{array}{ccccccc}
\mu & 0 & \cdots & 0 & 0 & \cdots & 0\\
0 & \nu & \cdots & 0 & 0 & \cdots & 0\\
\vdots & \vdots & & \vdots & \vdots & & \vdots\\
0 & 0 & \cdots & \nu & 0 & \cdots & 0\\
0 & 0 & \cdots & 0 & \lambda & \cdots & 0\\
\vdots & \vdots & & \vdots & \vdots & & \vdots\\
0 & 0 & \cdots & 0 & 0 & \cdots &  \lambda
\end{array}\right)
\]
where the eigenvalue $\nu$ appears $d$ times and the eigenvalue
$\lambda$ appears $d-1$ times.  By a simple phase redefinition we can
choose $\nu = 1$.  We will furthermore set $\lambda = -1$, which as we
shall see will give rise to a \Schrodinger\ equation in the continuum
limit.  As we shall discuss later, however, any value of $\lambda \neq
\mu$ gives a \Schrodinger\ equation; we use the $\lambda = -1$
condition merely to simplify the presentation.

With the stated conditions on $D$, we can compute $S$.  We find that all
elements of $S$ are equal to $\frac{1 + \mu}{2 d}$, except the matrix
elements $S_{i, i + d}$ and $S_{i + d, i}$, which are equal to
$\frac{1 + \mu}{2 d}-1$.  Thus,
\[
S_{ij} = \frac{1 + \mu}{2 d} -\delta_{0, | i-j | -d}
\]
At a microscopic level, this collision matrix gives an equal amplitude
for a particle to move in every direction other than directly
backwards.  To check the unitarity condition, we verify
\[
(2 d-1) \frac{(1 + \mu) (1 + \mu^*)}{4d^2}  +
 \frac{(1 + \mu -2 d) (1 + \mu^* -2 d)}{4d^2}
= 1.
\]
and
\[
(2d-2)  \frac{(1 + \mu) (1 + \mu^*)}{4d^2}  +
 \frac{(1 + \mu) (1 + \mu^* -2 d)}{4d^2}+
 \frac{(1 + \mu -2 d) (1 + \mu^*)}{4d^2}= 0.
\]

We can now proceed to calculate the other matrices needed for the
dynamics.  There are $d$ matrices $B^\alpha$.  For a particular value of
$\alpha$, we find that all matrix entries vanish except those in the
$(\alpha + 1)$th row and the $(\alpha + 1)$th column, which are given by
\[
(B^\alpha)_{\alpha + 1, i} =(B^\alpha)_{i,\alpha + 1} =
\left(\frac{1}{ \sqrt{d}} , 0^{d + \alpha -2}, 
-\frac{\sqrt{\alpha -1}}{\sqrt{\alpha}} ,
r (\alpha), r (\alpha + 1), \ldots, r (d-1)
\right)_i,
\]
where by $0^{k}$ we denote  a sequence of $k$ 0's, and where we have
defined
\[
r (\alpha) = \frac{1}{ \sqrt{\alpha (\alpha +1)}}.
\]
We can now immediately compute $G^\alpha$, which has nonzero elements
in the same positions, given by
\[
(G^\alpha)_{\alpha + 1, i} =
\left(\frac{-1}{(1 -\mu^*)\sqrt{d}} , 0^{d + \alpha -2},
\frac{\sqrt{\alpha -1}}{2\sqrt{\alpha}} , 
-\frac{1}{2} r (\alpha), -\frac{1}{2} r (\alpha + 1), \ldots,
-\frac{1}{2} r (d-1)
\right)_i
\]
\[
(G^\alpha)_{i,\alpha + 1} =
\left(\frac{-1}{(1 -\mu)\sqrt{d}}, 0^{d + \alpha -2},
\frac{\sqrt{\alpha -1}}{2\sqrt{\alpha}},
-\frac{1}{2} r (\alpha), -\frac{1}{2} r (\alpha + 1), \ldots,
-\frac{1}{2} r (d-1)
\right)_i.
\]
We are now interested in computing the differential equation describing
the continuum limit of the dynamics of the first component of $\zeta$,
which we will denote by $\Psi$.  As before, we define
\[
\Psi (\bfx, t) = \zeta_1 = \frac{\mu^{-\tau}}{ \sqrt{2 d}}  \left( \sum_{i}
\psi_i (\bfx, t) \right) +{\cal O} (\epsilon).
\]
To compute the dynamics of $\Psi$, we need to know only the first rows
of the matrices $B^\alpha G^\beta$ and $XC^\alpha C^\beta X^{-1}$.
{}From the above expressions, we find that the first row of $B^\alpha
G^\beta$ is given by
\[
(B^\alpha G^\beta)_{0i} = \frac{1}{  \sqrt{d}}  \delta^{\alpha \beta}
\left( -\frac{1}{\sqrt{d} (1 -\mu^*)} , 0^{d + \alpha -2},
\frac{\sqrt{\alpha -1}}{2\sqrt{\alpha}},-\frac{1}{2} r
(\alpha),-\frac{1}{2} r (\alpha + 1), \ldots, -r 
(d-1)
\right)_{i}
\]
{}From the above form of $X$, we see that the first row of
$XC^\alpha C^\beta X^{-1}$ is given by
\[
(XC^\alpha C^\beta X^{-1})_{0i} = \delta^{\alpha \beta}
\frac{1}{\sqrt{d}} \left( \frac{1}{\sqrt{d}}, 0^{d + \alpha -2}, -\frac{
\sqrt{\alpha -1}}{\sqrt{\alpha}},r (\alpha),r (\alpha + 1), \ldots,
r (d-1) \right)_{i}
\]
Thus, the first row of the combined matrix is
\[
(-B^\alpha G^\beta -\frac{1}{2} XC^\alpha C^\beta X^{-1})_{0i} =
\left(i \frac{1}{2 m} , 0^{2 d-1}  \right)_i
\]
where
\begin{equation}
m =d (\cot \theta -\csc \theta)
\label{eq:m}
\end{equation}
with
\begin{equation}
\mu = \cos \theta + i \sin \theta.
\label{eq:m2}
\end{equation}
As a result, we obtain the differential equation describing the
dynamical evolution of $\Psi$ in the continuum limit,
\begin{equation}
\partial_t {\Psi}(\bfx, t)
  = i \frac{1}{2 m}  \sum_{\alpha}\partial_\alpha^2 \Psi(\bfx, t),
\label{eq:Schrodinger}
\end{equation}
which we recognize as \Schrodinger\ evolution in $d$ dimensions.  In
Appendices \ref{sec:a1} and \ref{sec:a2} we work out the specific cases
of $d=2$ and $d=3$ in detail.

Note that had we chosen the eigenvalue $\lambda$ differently, the
difference would have appeared in the last $d-1$ rows and columns of
$G$.  Following through the computation, we find that the only change
would be that new terms would appear on the right hand side of
Eq.~(\ref{eq:Schrodinger}), proportional to the eigenvectors of $S$ with
eigenvalue $ \lambda$.  However, these terms would have had a phase
$(\mu \lambda^*)^\tau$, and thus would have averaged out in the
continuum limit as long as $\mu \neq \lambda$, making no change to the
final result Eq.~(\ref{eq:Schrodinger}).  Thus, we reach the conclusion
that for any collision rule invariant under the lattice symmetry group,
so long as $\mu$ is distinct from the other eigenvalues of the collision
matrix, the resulting continuum dynamics for the total amplitude $\Psi$
are governed by a \Schrodinger\ equation.

\subsection{Inclusion of a potential}

In general, we can easily include  an arbitrary potential $V (\bfx)$
by including a position dependent phase in the transition matrix $S$.
If we perform the above analysis for a model with transition matrix
\[
\tilde{S} (\bfx) = \exp (-i \epsilon^2 V (\bfx)) S
\]
where $S$ is a spatially invariant matrix such as discussed above,
then the general form of the dynamical equation becomes
\[
\partial_t {\eta}(\bfx, t)
 = - D^{-\tau} B^\alpha G^\beta D^\tau \partial_\alpha
\partial_\beta \eta(\bfx, t)
- \frac{1}{2} D^{-\tau} (X C^{\alpha} C^{\beta} 
X^{-1}) D^\tau \partial_\alpha
\partial_\beta \eta(\bfx, t)
-iV (\bfx) \eta (\bfx, t).
\]
This becomes for the total amplitude $\Psi$, the \Schrodinger\ equation
in an external potential
\[
\partial_t {\Psi}(\bfx, t)
  = i \frac{1}{2 m}  \sum_{\alpha}\partial_\alpha^2 \Psi(\bfx, t)
-iV (\bfx, t) \Psi(\bfx, t)
\]

\section{Many particles: Quantum lattice gas automata}
\label{sec:gas}

We now consider models in which multiple particles move
independently according to the \Schrodinger\ equation in $d$
dimensions.  One way of simulating the motion of $n$ particles
in $d$ dimensions is to introduce extra degrees of freedom for each
particle.  Thus, for example, we could model the motion of 2 particles
in one dimension by the lattice-Boltzmann model
\begin{equation}
\psi_{ik}  ( x + \epsilon c_i, y +  \epsilon c_k, t) = 
S_{il} S_{kj} \psi_{lj} (x, y, t -1)
\label{eq:boltzmann2}
\end{equation}
where $x,y$ are the positions of the two particles, $i, k$ are the
internal indices specifying their directions, and $S$ is a 2-by-2 matrix
for unitary \Schrodinger\ evolution in one dimension, as discussed in
Sec.~\ref{sec:se1d}.  Notice that this dynamics is equivalent to that of
a single particle moving in two dimensions.

In a similar fashion we can describe models where $n$ particles
move in $d$ dimensions, by constructing a
unitary lattice-Boltzmann model in $nd$ dimensions.  It is
straightforward to incorporate an arbitrary interparticle potential in
this formulation; the potential is a function of the particle
positions and can be included as discussed in Section III.C.

This gives a procedure for simulating an interacting nonrelativistic
quantum many-body system on a classical computer.  Although this may
give a useful algorithm for systems containing only a few particles,
if we wish to simulate the motion of a large number of particles using
the method just described it is clear that the number of calculations
needed to perform even one time step of the evolution become rapidly
intractable.  For example, simulating the motion of $20$ particles in
three dimensions on a lattice of side length $100$ would take on the
order of $10^{120}$ calculations per time step, beyond the capacity of
any imaginable classical computer.

However, the technology of {\it quantum
  computing}\cite{bib:quantumcomputing} presents a paradigm in which
such calculations can be done.  We will now describe a way in which the
above algorithm can be implemented on a quantum computer with a speedup
exponential in the number of particles.  In fact, it is natural to
simultaneously perform the calculation for all numbers of particles
which will fit on the lattice, essentially performing a discrete
simulation of nonrelativistic quantum many-body theory.  The resulting
model falls in the class of quantum lattice gas automata, which were
recently defined by Meyer~\cite{bib:meyer} in the context of the $(1 +
1)$-dimensional Dirac model.  The exponential speedup of this algorithm
on a quantum computer is a specific example of the general observation
by Feynman~\cite{bib:feynman} and Lloyd~\cite{bib:lloyd} that quantum
mechanical systems can be simulated more efficiently on a quantum
computer than on a classical computer.

A quantum-computing device is composed of simple quantum elements such
as particles with spin $1/2$ (q-bits).  The state space of the system
at any fixed time is the tensor product of the Hilbert spaces of the
states of the elementary computational elements.  Thus, for example, a
system with $m$ q-bits has a state space of dimension $2^m$.  At each
time step, some small number of q-bits (usually two or
three~\cite{bib:quantumgate}) are subjected to a unitary time
evolution, described by acting with a unitary matrix on the Hilbert
space of the affected elements.  Quantum computers have recently
become of great interest because of the result due to
Shor\cite{bib:Shor} that it is possible to factor large integers on a
quantum computer in polynomial time, a procedure thought to be
impossible on a classical computer.

In order to implement the many-body simulation described in the
beginning of this section on a quantum computer, it is necessary to make
some restrictions on the behavior of the many-body wavefunction under
exchange of particles.  The example system in Eq.~(\ref{eq:boltzmann2})
describes two particles moving in one dimension without interacting.  In
this model, both particles can be moving in the same direction from the
same lattice site at a given point in time.  We can modify this model
slightly to give the particles exclusionary (Fermi) statistics by making
the transition matrix at $x =y$ force the two particles to move in
different directions.  This corresponds to introducing a contact
interaction between the two particles when they move within a single
lattice distance.  By making the initial conditions antisymmetric under
exchange of $x$ and $y$, we have a simulation of two nonrelativistic
fermions moving in one dimension.  Alternatively, we could symmetrize
the wavefunction and we would have a simulation of ``hard bosons'' which
obey Bose statistics but which cannot occupy the same lattice site.
Either of these approaches naturally generalize to arbitrary numbers of
particles and arbitrary dimensions.  For the remainder of the discussion
we assume that the particles obey Bose statistics.  The issue of
implementing fermionic systems on a quantum computer is more
subtle~\cite{bib:feynman}, and has recently been addressed by Abrams and
Lloyd~\cite{bib:al}.

In previous sections we discussed the motion of a single particle,
with a wave function $\psi_k (\bfx, t)$.  Now, we would like to
consider the state space for a quantum system of many particles.  A
natural basis for the Hilbert space of such a system is the set of
states in the fermionic Fock space associated with the spatial
lattice; such states are identified by a set of occupation numbers
$s_k (\bfx)$ (taking values 0 or 1) for each possible particle
position $\bfx$ and internal index $k$.  The Hilbert space of the
model is thus $2^{ml^d}$ dimensional, where $m$ is the number of
possible values of the internal index, and $l^d$ is the number of
lattice sites.  For example, a basis vector of the state space for a
one-dimensional system with 4 lattice sites $x = 1, 2, 3, 4$ might be
given by
\begin{equation}
| s \rangle = |(s_2 (1), s_1 (1)), \ldots,(s_2 ( 4), s_1 (  4))
\rangle =
| (0,1),(0, 0), (0, 0),(1,1) \rangle
\label{eq:simplestate}
\end{equation}
where each ordered pair corresponds to the occupation numbers at a
given lattice site.  Thus, this state corresponds to the configuration
where a single particle is at $x = 1$, with $k = 1$, and both particle
positions at $x = 4$ are filled.  

The state of the quantum system at any given value of the
discrete time parameter $t$ is given by a vector
\[
| \psi (t)\rangle = \sum_{s} C_s (t) | s \rangle
\]
where the sum is taken over all basis vectors of the Hilbert space.
This state is defined by the coefficients $C_s (t)$.  In the quantum
computing paradigm, this corresponds to the state space of $m l^d$
independent q-bits.

We will now define a quantum lattice gas automaton by defining a
dynamics on the quantum state space.  The dynamics of the quantum
lattice-gas will be described in two steps, just as in classical
lattice-gas automaton models.  First there is a collision step in
which the particles at each lattice site interact.  Then there is an
advection step, describing the propagation of the particles in the
directions associated with the vectors $\bfc_k$.  Each of these steps
is described in the quantum system by a unitary transfer matrix acting
on the state space of the system.  The total dynamics can then be
described by the equation
\[
| \psi (t + \Delta t) \rangle = A \cdot K \cdot | \psi (t) \rangle
\]
The advection step simply corresponds to a permutation matrix $A$ on the
basis vectors described above, where each bit is moved forward in the
direction corresponding to the appropriate vector $\bfc_k$.  For
example, acting on the state in Eq.~(\ref{eq:simplestate}), the result
of applying the advection operator would be
\[
A | s \rangle =| (0,1),( 0,1), (1, 0),  (0, 0) \rangle
\]
where we assume periodic boundary conditions on the lattice.  The
particle which was at lattice site $x = 1$ has been advected to $x =
2, k = 1$, and the two particles which were at $x = 4$ have moved to
$x = 3$ and $x = 1$.

We now consider the collision part of the time development rule.  The
collision process is defined by a single unitary $2^m$ by $2^m$ matrix
$T$, which acts separately on the quantum bits associated with each
lattice site.  Thus, the state of the system is transformed by the
unitary matrix
\[
K = T \otimes T \otimes \cdots \otimes T
\]
given by the $l^d$-fold tensor product of $T$.  We would like the
collision matrix $T$ to have the property that it conserves particle
number.  Thus, this matrix is block diagonal in the subspaces of the
Hilbert space at each lattice site corresponding to a fixed particle
number.

We have now defined a discrete model for quantum many-particle
systems.  To understand the behavior of this model in the continuum
limit, let us consider the behavior when the number of particles in
the system is relatively small compared to the number of lattice
sites.  In this case, at most lattice sites the number of particles
present will be either 0 or 1.  This part of the dynamics, which
describes the free propagation of single particles, is described by
the part of $T$ in the single-particle Hilbert space.  However,
because this is a unitary matrix, generically the dynamics described
by this transition matrix is precisely that which we studied in the
previous sections, and corresponds to a nonrelativistic particle
propagating according to the \Schrodinger\ equation.  Thus, for
relatively sparse systems, this quantum lattice-gas model simulates a
system of many nonrelativistic particles whose free propagation is
given by the \Schrodinger\ equation.  The remaining parts of the
collision matrix $T$ describe a contact interaction between the
various particles.

Let us now discuss the computational complexity of the quantum
algorithm.  To implement the advection transformation by using quantum
computing elements, it is only necessary to perform a series of
exchanges of the values of the quantum bits representing the particle
occupation numbers.  The number of such exchanges is essentially equal
to the number of bits, $m l^d$ (recall that on a Euclidean lattice of
dimension $d$, $m = 2d$, so that for example if $d = 3$, the advection
operation can be implemented in approximately $6l^3$ quantum
operations).

The matrix $T$ acts on the Hilbert space associated with a subset of
$m$ of the q-bits in the system.  Counting degrees of freedom,
generically such a matrix can be implemented with approximately
$2^{2m}/15$ elementary quantum operations on pairs of q-bits.  For
example, in a 3D system, it would take on the order of 300 quantum
operations to implement each $T$ matrix, so that the number of
computational steps needed to perform the transformation by $K$ would
be around $300 l^3$.  Note, however, that the part of $T$ which acts
on the multiple particle Hilbert space simply changes the phases of a
delta function type interaction between the particles.  Since these
phases may not affect the results in most problems of physical
interest, these components of $T$ can be arbitrary.  Thus, in practice
we need only find a combination of operations on q-bits which will
give a matrix $T$ which preserves particle number and gives the
desired symmetry properties and eigenvalues, reducing the number of
steps needed significantly below 300.

Combining these observations, we see that
this model can be simulated with on the order of $l^d$
elementary quantum computations at each time step (or on the order of
1 if we are using a quantum computing system which allows parallel
computation).  Since this system automatically contains the
multi-particle wave function for all possible particle numbers, we
have achieved an exponential increase in speed over what was possible
on a classical computer.

The system as defined so far only includes interactions
between the particles in the form of delta function interactions
parameterized by the components of $T$ in the multiple particle space.
We can introduce an arbitrary interparticle potential $V (\bfx,\bfy)$
by hand, by multiplying the wave function at each time step by the
tensor product over all pairs of q-bits
\[
U = \otimes U_{i,j,\bfx,\bfy}
\]
where the matrix
\[
U_{i,j,\bfx,\bfy}= \left(
\begin{array}{cccc}
1 & 0 & 0 & 0\\
0 & 1 & 0 & 0\\
0 & 0 & 1 & 0\\
0 & 0 & 0 &  e^{-i \epsilon^2 V (\bfx,\bfy)}
\end{array} \right)
\]
acts on the Hilbert space associated with the q-bits $s_i (\bfx)$ and
$s_j (\bfy)$, changing the phase of the wavefunction only in the
component where both q-bits have the value 1.  Implementing this
interparticle potential will take on the order of $m^2 l^{2d}$ quantum
computations for each time step.  Although this significantly increases
the computational complexity of the quantum algorithm, this is still
exponentially faster than the analogous classical algorithm, since the
particle number $n$ does not affect the complexity.  Note that, unlike
the rest of the algorithm, the implementation of interparticle
potentials involves nonlocal interactions on the lattice.

To clarify the discussion, we consider a simple example of a collision
matrix $T$.  For a many-body system in one dimension, at each lattice
site the collision matrix $T$ is a 4-by-4 matrix, acting on the
Hilbert space with basis $| (0,0) \rangle$, $| (0,1) \rangle$, $|
(1,0) \rangle$, $| (1,1) \rangle$.  Since we are assuming that
particle number is conserved and that the dynamics is symmetric under
left-right reflection, the matrix $T$ is of the form
\[
T = \left(\begin{array}{cccc}
\alpha& 0 & 0 & 0\\
0 &a& b & 0\\
0 &b &a& 0\\
0 &0 &0 & \beta
\end{array}\right)
\]
where $\alpha,\beta,a,b$ are complex numbers satisfying $| \alpha |^2 =
| \beta |^2 = | a |^2 + | b |^2 = 1$ and $a \bar{b} + \bar{a}b = 0$.  By
a simple global phase redefinition, we can choose $\alpha = 1$.  The
part of $T$ in the single-particle Hilbert space is precisely the form
of the collision matrix $S$ from Sec.~\ref{sec:se1d}.  {}From the
eigenvalues of this matrix we can determine the mass of the free
particles in the model.  Finally, there is a single parameter $\beta$
which describes the phase with which two particles ``bounce''.  In this
simple one-dimensional model, there is therefore little freedom in
choosing the particle interaction.  In higher dimensions there would be
nontrivial phases describing delta function interactions between up to
$2d$ particles.

One major concern in the implementation of any algorithm is the issue
of precision.  This problem is particularly acute on a quantum
computer, where each quantum operation involves acting on the state
with a unitary transformation which can only be controlled up to some
finite precision.  Furthermore, on a quantum computer there is the
related but distinct problem of decoherence which must be addressed in
order for any quantum computation to be feasible.  There has been a
great deal of work recently describing how these problems can be
solved using dynamical quantum error correction
methods~\cite{bib:tolerant}.  Without going into this issue in depth,
we make the simple observation that even without error correction, if
the precision of each quantum operation is $1-1/t$, then at each time
step the error in the wave function will take a random step in the
Hilbert space with size $1/t$.  Only after on the order of $t^2$
operations will this error become significant.  Thus, if we could
achieve a precision better than $10^{-5}$, we could perform $10^{10}$
quantum operations successfully, which would allow us to simulate for
example an interacting 3D system on a lattice with size of order
$20^3$.  With the error correction schemes described
in~\cite{bib:tolerant}, there is in principle no upper bound on how
large a system could be simulated, other than the size of the quantum
computer which could be built to perform the simulation.

Finally, we consider the issue of measurement in quantum lattice
gases.  In a classical lattice gas, hydrodynamic quantities, such as
mass and momentum density, are obtained by averaging particles' mass
and momentum over blocks in space and/or time.  In a typical
lattice-gas simulation, this is done from time to time to obtain the
macroscopic variables of interest.  The process of measuring these
quantities is purely passive -- that is, their measurement does not
affect the subsequent dynamical evolution at all.  In contrast, the
analogous operation for a quantum lattice gas would involve
occasionally measuring the state of some subset of the q-bits in the
system, thus collapsing the quantum wavefunction onto the eigenstates
of the (space and/or time) block number operator.  The set of
quantities which are accessible through this type of simulation are
rather different from those accessible through simulation methods on a
classical computer.  For example, the dynamics of the system defines
an effective Hamiltonian which is an approximation to the Hamiltonian
of the many-body quantum system being simulated, however the spectrum
of this Hamiltonian is not directly amenable to measurement.  Instead,
the types of observables which can be measured in the simulation are
precisely equivalent to the types of observables which can be measured
in an actual interacting quantum system.  For example, a typical
experiment might be to initialize the system in a particular known
state at time $t = 0$, and to ask for the probability $p$ at time $t$ that
there is a particle in a region of space $dx^3$.  Just as in the
physical quantum system, we can ask such a question of our simulation;
we can perform the experiment a number of times, and each time we will
find a particle with probability $p$.  To actually compute $p$ to some
degree of accuracy requires repeating the experiment a number of times.

\section{Numerical Results}
\label{sec:numeric}

To test the algorithm, we consider the dispersion relation of plane
waves in periodic geometry in two dimensions.  We consider a periodic
grid with dimensions $N$ by $N$, and initialize it with a plane wave
of the form
\[
\psi_j(\bfx, 0) =
  \frac{1}{4}\exp\left(i{\bf k}\cdot\bfx - i\omega t\right)
\]
for $j=1,\ldots,4$, where
\[
{\bf k} = 2\pi\left(l_x\hat{\bfx} +
                    l_y\hat{\bfy}\right),
\]
where $l_x$ and $l_y$ are
integers.  Choosing units where the spatial dimensions are of unit
length, we have $\epsilon = 1/N$, and $\Delta t = \epsilon^2 = 1/N^2$

We evolve this initial condition in time, using Eq.~(\ref{eq:tdcm}) with
$\mu=-i$ (hence $m=2$) for the collisions.  Every four time steps, we
measure the inner product of the wave function with its initial
condition,
\[
S(t)\equiv
 \frac{1}{N^2}\sum_\bfx
 \Psi^* (\bfx, 0) \Psi (\bfx, t).
\]
The result should go like $\exp(-i\omega t)$, so the ratio of two
successive values of this quantity is
\[
\frac{S(t+ 4\dt)}{S(t)}=\exp(-4i\omega\dt),
\]
and hence the frequency is given by
\[
\omega = \frac{i}{4\dt}\ln\left(\frac{S(t+4\dt)}{S(t)}\right).
\]
For a given wavevector ${\bf k}$, we measure this frequency at many time
steps $t$ and take an average.

We expect the evolution of the system to be governed by the
\Schrodinger\ equation,
\[
\partial_t\Psi = \frac{i}{2m}\partial^2\Psi,
\]
Since $m=2$, this leads to the
dispersion relation
\[
\omega = \frac{k^2}{4} .
\]

\begin{figure}
\begin{center}
\leavevmode
\hbox{%
\epsfxsize=3.5in
\epsffile{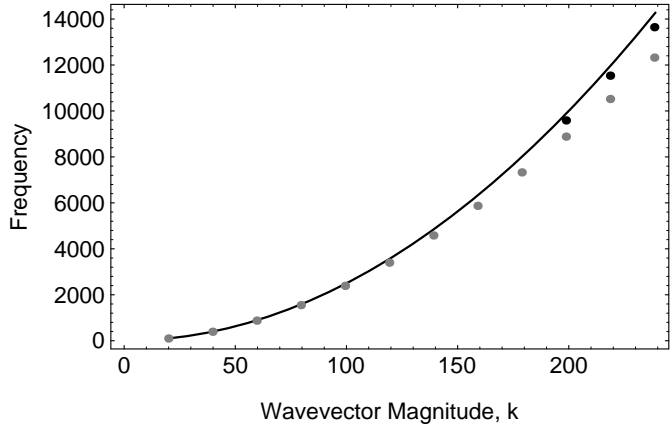}}
\end{center}
\caption{{\bf Plane-Wave Dispersion Relation} is shown for $N_x=256$
  (gray points) and $N_x=512$ (black points).}
\label{fig:disp}
\end{figure}

We performed a series of simulations, where we considered wavenumbers
$l_x=3l$ and $l_y=l$, where $l\in\{1,\ldots,12\}$.  The points plotted
in Fig.~\ref{fig:disp} show the measured frequency $\omega$ as a
function of $|{\bf k}|=2 \pi\sqrt{l_x^2+l_y^2}$.  The solid curve is
$|{\bf k}|^2/4$.  It is evident that the agreement is excellent in the
``hydrodynamic'' limit of small $|{\bf k}|$, but degrades due to lattice
artifacts of order $|{\bf k}|\Delta x$ at higher wavevector magnitudes.
To demonstrate this, we include data for $N=256$ (gray points) and
$N=512$ (black points).  It is evident that the dispersion relation is
valid for higher wavenumbers on the larger lattice.

\section{Conclusions}

We have considered a very general class of lattice models satisfying
unitary time-evolution rules.  We have shown that generic models of
this type describe the evolution of a nonrelativistic particle
according to the \Schrodinger\ equation in an arbitrary number of
dimensions.  These models can naturally be used to construct quantum
lattice gases describing nonrelativistic many-body physics in an
arbitrary number of dimensions.  It is straightforward to include
an arbitrary interparticle potential into these models.

There are many ways in which this work could be extended.  Numerical
simulations could be performed in an arbitrary number of dimensions with
multiple particles and with nontrivial spatially dependent potentials,
and the results checked in analytically tractable cases.  Further
analysis is needed to understand the behavior of the system in the
regime where particles are dense.  It might also be interesting to
consider more general collision rules which create and destroy
particles, possibly including antiparticles with separate quantum
numbers.

Of course, the actual implementation of quantum lattice-gas models on
quantum computing devices is something which may not be possible for
many years, if ever.  However, these lattice models give a simple
framework with which to study problems in many-body theory.
Furthermore, the methods described here are also quite practical for
simulating systems of a few particles on a classical computer.  It may
be that the exact unitarity of these models at a microscopic level
will make them more stable and possibly more useful than currently
used discrete methods such as finite-difference approaches.

\section*{Acknowledgements}
We would like to acknowledge helpful conversations with Francis
Alexander, Peter Coveney, Eddie Farhi, and Jeffrey Yepez.  BMB was
supported in part by Phillips Laboratories and by the United States Air
Force Office of Scientific Research under grant number F49620-95-1-0285.
WT was supported in part by the divisions of Applied Mathematics of the
U.S.  Department of Energy (DOE) under contracts DE-FG02-88ER25065 and
DE-FG02-88ER25066, in part by the U.S. Department of Energy (DOE) under
cooperative agreement DE-FC02-94ER40818, and in part by the National
Science Foundation (NSF) under contract PHY90-21984.

\appendix

\section{Schr\"{o}dinger equation in 2D}
\label{sec:a1}

We now present the formalism described above explicitly in
two dimensions.  The matrices $D$ and $X$ are given by
\[
D = \left(\begin{array}{cccc}
\mu& 0 & 0 & 0\\
0 &1& 0 & 0\\
0 &0 &1& 0\\
0 &0 &0 &-1
\end{array}\right)
\]
\[
X = \left(\begin{array}{cccc}
1/2 & 1/2 & 1/2 & 1/2\\
\frac{1}{ \sqrt{2}} & 0&-\frac{1}{ \sqrt{2}} & 0\\
0 &\frac{1}{ \sqrt{2}} & 0 &-\frac{1}{ \sqrt{2}} \\
1/2 & -1/2 & 1/2 & -1/2
\end{array} \right).
\]
This gives us the collision matrix
\begin{equation}
S =X^{-1} DX =\frac{1}{ 4} \left(\begin{array}{cccc}
\mu + 1 &\mu + 1 &\mu - 3 &\mu + 1 \\
\mu + 1 &\mu + 1 &\mu + 1 &\mu - 3 \\
\mu - 3 &\mu + 1 &\mu + 1 &\mu + 1 \\
\mu + 1 &\mu - 3 &\mu + 1 &\mu + 1
\end{array}
\right).
\label{eq:tdcm}
\end{equation}
We can compute
\[
B^1 =X C^1 X^{-1}= \left(\begin{array}{cccc}
0 &   \frac{1}{\sqrt{2}}  &  0 & 0\\
\frac{1}{\sqrt{2}}  &  0 &   0 &\frac{1}{\sqrt{2}}\\
0 &  0 &  0 & 0\\
0 &  \frac{1}{\sqrt{2}} &  0 & 0
\end{array} \right)
\]
\[
B^2 =X C^2 X^{-1}= \left(\begin{array}{cccc}
0 &  0 &  \frac{1}{\sqrt{2}} & 0\\
0 &  0 &  0 & 0\\
\frac{1}{\sqrt{2}} &  0 &  0 & -\frac{1}{\sqrt{2}}\\
0 & 0 & - \frac{1}{\sqrt{2}} &  0 
\end{array} \right),
\]
and thus
\[
G^1 =X C^1 X^{-1}= \left(\begin{array}{cccc}
0 & -  \frac{1}{(1 -\mu)\sqrt{2}}  &  0 & 0\\
-\frac{1}{(1 -\mu^*)\sqrt{2}}  &  0 &   0 &-\frac{1}{2\sqrt{2}}\\
0 &  0 &  0 & 0\\
0 &  -\frac{1}{2\sqrt{2}} &  0 & 0
\end{array} \right)
\]
\[
G^2 =X C^2 X^{-1}= \left(\begin{array}{cccc}
0 &  0 &  -\frac{1}{(1 -\mu)\sqrt{2}} & 0\\
0 &  0 &  0 & 0\\
-\frac{1}{(1 -\mu^*)\sqrt{2}} &  0 &  0 & \frac{1}{2\sqrt{2}}\\
0 & 0 &  \frac{1}{2\sqrt{2}} &  0 
\end{array} \right).
\]
Combining these matrices together we find
\[
\partial_t {\zeta} =
\left(\begin{array}{cccc}
\frac{i}{2m} & 0 & 0 & 0\\
0 &-\frac{i}{2m} & 0 & 0\\
0 &0 &0& 0\\
\frac{i (-\mu)^\tau}{2m}  &0 &0 & 0
\end{array}\right) \partial_x^2 \zeta
 +\left(\begin{array}{cccc}
\frac{i}{2m} & 0 & 0 & 0\\
0 &0 & 0 & 0\\
0 &0 &-\frac{i}{2m}& 0\\
\frac{i (-\mu)^\tau}{2m}  &0 &0 & 0
\end{array}\right)  \partial_y^2 \zeta
+\left(\begin{array}{cccc}
0& 0 & 0 & 0\\
0 &0& -\frac{i}{2m} & 0\\
0 &-\frac{i}{2m} &0& 0\\
0 &0 &0 & 0
\end{array}\right) \partial_x\partial_y \zeta
\]
with $m$ described as in Eqs.~(\ref{eq:m}) and (\ref{eq:m2}), with $d =
2$.

As predicted by the general discussion above, the total amplitude
contained in the first component of $\zeta$ satisfies a \Schrodinger\
equation
\[
\partial_t {\Psi}(\bfx, t)
 = i \frac{1}{2 m}  (\partial_x^2 + \partial_y^2) \Psi(\bfx, t)
\]
where
\[
\Psi (\bfx, t) = \frac{\mu^{-\tau}}{ 2} \left[ \psi_1 (\bfx, t)+\psi_2
(\bfx, t) +\psi_3 (\bfx, t)+\psi_4 (\bfx, t) \right].
\]
It is interesting to note that while the variation of the fourth
component of $\zeta$ contains an oscillating phase, and thus has no
interesting behavior on the time scale of interest, the second and third
components obey separate second-order differential equations analogous
to the \Schrodinger\ equation, but without rotational invariance.

\section{Schr\"{o}dinger equation in 3D}
\label{sec:a2}

Using the above formalism in 3D, we have
\[ X =
\left(\begin{array}{cccccc}
 {1\over {{\sqrt{6}}}}&{1\over {{\sqrt{6}}}}&{1\over {{\sqrt{6}}}}&
   {1\over {{\sqrt{6}}}}&{1\over {{\sqrt{6}}}}&{1\over {{\sqrt{6}}}}\\
   {1\over {{\sqrt{2}}}}&0&0&-{1\over {{\sqrt{2}}}}&0&0\\
   0&{1\over {{\sqrt{2}}}}&0&0&-{1\over {{\sqrt{2}}}}&0\\
   0&0&{1\over {{\sqrt{2}}}}&0&0&-{1\over {{\sqrt{2}}}}\\
   {1\over 2}&-{1\over 2}&0&{1\over 2}&-{1\over 2}&0\\
   {1\over {2\,{\sqrt{3}}}}&{1\over {2\,{\sqrt{3}}}}&-{1\over {{\sqrt{3}}}}&
   {1\over {2\,{\sqrt{3}}}}&{1\over {2\,{\sqrt{3}}}}&-{1\over {{\sqrt{3}}}}
\end{array}\right) 
\]
\[ D =
\left(\begin{array}{cccccc}
\mu&0&0&0&0&0\\ 0&1&0&0&0&0\\ 0&0&1&0&0&0\\ 0&0&0&1&0&0\\
   0&0&0&0&-1&0\\ 0&0&0&0&0&-1
\end{array}\right) 
\;\;\;\;\; S = \frac{1}{6} 
\left(\begin{array}{cccccc}
 \mu + 1&\mu + 1&\mu + 1&\mu -5&\mu + 1&\mu + 1\\
   \mu + 1&\mu + 1&\mu + 1&\mu + 1&\mu -5&\mu + 1\\
   \mu + 1&\mu + 1&\mu + 1&\mu + 1&\mu + 1&\mu -5\\
   \mu -5&\mu + 1&\mu + 1&\mu + 1&\mu + 1&\mu + 1\\
   \mu + 1&\mu -5&\mu + 1&\mu + 1&\mu + 1&\mu + 1\\
   \mu + 1&\mu + 1&\mu -5&\mu + 1&\mu + 1&\mu + 1
\end{array}\right) 
\]
\[ B^1 =
\left(\begin{array}{cccccc}
 0&{1\over {{\sqrt{3}}}}&0&0&0&0\\
   {1\over {{\sqrt{3}}}}&0&0&0&{1\over {{\sqrt{2}}}}&
   {1\over {{\sqrt{6}}}}\\ 0&0&0&0&0&0\\ 0&0&0&0&0&0\\
   0&{1\over {{\sqrt{2}}}}&0&0&0&0\\
   0&{1\over {{\sqrt{6}}}}&0&0&0&0
\end{array}\right) 
\;\;\;\;\;
 G^1 =
\left(\begin{array}{cccccc}
 0&-{1\over {{\sqrt{3}}\,\left( 1-\mu \right) }}&0&0&0&0\\
   -{1\over {{\sqrt{3}}\,\left( 1-\mu^* \right) }}&0&0&0&
   {{-1}\over {2\,{\sqrt{2}}}}&{{-1}\over {2\,{\sqrt{6}}}}\\
   0&0&0&0&0&0\\ 0&0&0&0&0&0\\
   0&{{-1}\over {2\,{\sqrt{2}}}}&0&0&0&0\\
   0&{{-1}\over {2\,{\sqrt{6}}}}&0&0&0&0
\end{array}\right) 
\]
\[ B^2 =
\left(\begin{array}{cccccc}
 0&0&{1\over {{\sqrt{3}}}}&0&0&0\\ 0&0&0&0&0&0\\
   {1\over {{\sqrt{3}}}}&0&0&0&-{1\over {{\sqrt{2}}}}&
   {1\over {{\sqrt{6}}}}\\ 0&0&0&0&0&0\\
   0&0&-{1\over {{\sqrt{2}}}}&0&0&0\\
   0&0&{1\over {{\sqrt{6}}}}&0&0&0
\end{array}\right) 
\;\;\;\;\;G^2 = \left(\begin{array}{cccccc}
 0&0& -{1\over {{\sqrt{3}}\,\left( 1-\mu \right) }}&0&0&0\\
   0&0&0&0&0&0\\  -{1\over {{\sqrt{3}}\,\left( 1-\mu^* \right) }}&
   0&0&0&{1\over {2\,{\sqrt{2}}}}&{{-1}\over {2\,{\sqrt{6}}}}\\
   0&0&0&0&0&0\\ 0&0&{1\over {2\,{\sqrt{2}}}}&0&0&0\\
   0&0&{{-1}\over {2\,{\sqrt{6}}}}&0&0&0
\end{array}\right) 
\]
\[ B^3 =
\left(\begin{array}{cccccc}
 0&0&0&{1\over {{\sqrt{3}}}}&0&0\\ 0&0&0&0&0&0\\ 0&0&0&0&0&0\\
   {1\over {{\sqrt{3}}}}&0&0&0&0&-{\sqrt{{2\over 3}}}\\ 0&0&0&0&0&0\\
   0&0&0&-{\sqrt{{2\over 3}}}&0&0
\end{array}\right) 
\;\;\;\;\;
G^3 = \left(\begin{array}{cccccc}
0&0&0& -{1\over {{\sqrt{3}}\,\left( 1-\mu \right) }}&0&0\\
   0&0&0&0&0&0\\ 0&0&0&0&0&0\\
   -{1\over {{\sqrt{3}}\,\left( 1-\mu^* \right) }}&0&0&0&0&
   {1\over {{\sqrt{6}}}}\\ 0&0&0&0&0&0\\
   0&0&0&{1\over {{\sqrt{6}}}}&0&0
\end{array}\right) 
\]
{}From these matrices we find that the total amplitude
\[
\Psi (\bfx, t) = \frac{\mu^{-\tau}}{\sqrt{6}} \left[ \psi_1 (\bfx, t)+\psi_2
(\bfx, t) +\psi_3 (\bfx, t)+\psi_4 (\bfx, t) 
 +\psi_5 (\bfx, t)+\psi_6 (\bfx, t) 
\right].
\]
satisfies the \Schrodinger\ equation
\[
\partial_t {\Psi}(\bfx, t) = i \frac{1}{2 m}  (\partial_x^2 + \partial_y^2
+\partial_z^2) \Psi(\bfx, t)
\]
where as usual $m$ is related to $\mu$ through Eqs.~(\ref{eq:m}) and
(\ref{eq:m2}), with $d = 3$.

\end{document}